\documentclass{article}
\usepackage{amsmath}

\begin{document}
\title{ A String Motivated Approach to the Relativistic Point Particle.}
\author{Michael Tuite \\
{\em Department of Mathematical Physics,}\\
{\em National University of Ireland, Galway, Ireland} \\ \\
Siddhartha Sen\\
{\em School of Mathematics, Trinity College Dublin, Dublin2, Ireland}\\
{\em I.A.C.S., Jadavpur, Calcutta 700032, India.}}
\date{}
\maketitle

\begin{abstract}

Using concepts developed in string theory, Cohen, Moore, Nelson and
Polchinski calculated the propagator for a relativistic point particle.
Following these authors we extend the technique to include the case of
closed world lines.
The partition function found corresponds to the Feynmann and Schwinger
proper time formalism. We also explicitly verify that the partition
function is equivalent to the usual path length action partition function.
As an example of a sum over closed world lines, we compute the
Euler-Heisenberg effective Lagrangian in a novel way.

\end{abstract}

\section*{Introduction}
I am very happy to present this work on the occasion of Bal's birthday
celebrations. I hope the simple concrete calculations presented
give him some enjoyment. The work reported was done in collaboration
with Michael Tuite \cite{1}.
The quantization of the free relativistic point particle is perhaps the most
basic system with constraints studied in physics \cite{2}. In this paper we
follow
the string motivated technique developed by Cohen, Moore, Nelson and
Polchinski
\cite{3} for considering the bosonic point particle. The usual action is
proportional
to the world line path length and is analagous to the string Nambu-Goto
action
\cite{3}.

Alternatively, a world line metric can be introduced to obtain a more
tractable
expression in analogy to the Polyakov action \cite{3,4}. These two actions
are known to be equivalent at both the classical and quantum level
\cite{4,5}.
Considering the reparameterization invariance of the Polyakov-like action,
as in ref.
\cite{3}, the partition function can be reduced to a sum over embeddings and
a single
parameter. This parameter is analogous to the set of modular parameters
of a Riemann surface in string theory \cite{6,7}. The dependence of the
partition function on this parameter is shown to be different for sums
over open and closed particle world lines because of the presence of a
diffeomorphism zero mode in the latter case. The parameter in these two
cases plays the role of the ficticious "proper time" in the Feynmann
\cite{8} and
Schwinger \cite{8} normalization prescription that this partition function
is equivalent to the original path length action partition function. As an
example of a process involving closed world lines, we compute the
Euler-Heisenberg effective Lagrangian \cite{9} for a boson interacting with
a constant
external electromagnetic field.

\section*{The Relativistic Point Particle}

We begin by reviewing the definition of the free relativistic Euclidean
point particle Lagrangian (in $d$ dimensions) which is analogous to the
Nambu-Goto
Lagrangian of string theory \cite{4}. The action $S$ is proportional to the
path length
(proper time)
\[
S[x_\mu] = m \int^1_0(\dot{x}_\mu^2)^{\frac{1}{2}} dt,
\]
where $t$ is a parametrization of the path. In analogy to the Polyakov
string
we can introduce metric $g(t)$ along the world line and define an
alternative Polyakov action, $S_g$
\cite{3,4}
\begin{align}
S_g[x_\mu,g]&=\frac{1}{2}\int^1_0 \sqrt{g}(g^{-1} \dot{x}^2_\mu+m^2)dt \\
&= \frac{1}{2}\int_0^1(e^{-1}\dot{x}^2_\mu+m^2e)dt \\
\intertext{where $e=\sqrt{g}$ is the \" einbein \". It is straightforward to
show that }
S_g[x_\mu,e] &\geq S_g[x_\mu,\hat{e}]=S[x_\mu],\\
\intertext{where $\hat{e}$ is the induced einbein}
\hat{e} &=\frac{1}{m}(\dot{x}^2_\mu)^{\frac{1}{2}}
\end{align}
and hence (1) and (3) describe the same classical system. Alternatively,
solving for the equations of motion one finds the constraint $e=\hat{e}$.

Both $S$ and $S_g$ are reparametrization invariant under $t \to s(t)$ with
$\frac{ds}{dt}>0$ where
\begin{align}
\frac{d x_\mu(t)}{dt} &\to \frac{dt}{ds} \frac{d x_\mu(t(s))}{dt} \\
e(t) &\to \frac{dt}{ds} e(t(s)).
\end{align}
This transformation must however respect the boundary conditions on the
world
line. Thus for an open path $s(0)=0, s(1)=0$ whereas for a closed path
$s(t)=s(t+1)$. The parameter $c=\int e(t)dt$ remains invariant and can be
used to label diffeomorphically inequivalent metrics. It is analogous to the
set
of modular parameters of a Riemann surface in the Polyakov string formalism
\cite{6,7}.

The quantum theories are now defined by the partition functions
\begin{align}
Z &= \int [dx_\mu] \exp(-S) \\
Z_g &= \int [de][dx_\mu] \exp(-S_g)
\end{align}
We now exploit the reparametrization invariance of $S_g$ to extract a formal
diffeomorphic volume factor in (8). We change variables from $e(t)$
to $c,f(t)$ where $f(t)$ is the reparametrization which transforms $e(t)$
to $c$ \cite{3}. From (6) we find
\[
f'(t) e(f(t)) = c.
\]
The Jacobian $J$ for this change of variables is most conveniently computed
in the tangent space of einbeins $\{ \delta e \}$ \cite{3,6}. We find the
equivalent Jacobian for the transformation from $\delta e$ to $\delta c$,
$\zeta$ where
$\zeta(t)=\delta f(f'(t))$ is an infinitesimal diffeomorphism vector field
i.e.$[d(\delta e)]=Jd(\delta c)[d\zeta]$.

We define the normalization for the measure $[d(\delta e)]$ by
\[
\int [d(\delta e)] \exp(-\frac{1}{2}\mid\mid \delta e \mid\mid^2)=1,
\]
where the invariant norm is
\[
\mid\mid \delta e \mid\mid^2 = \int^1_0 e^{-1}\delta e^2 dt.
\]
 We find from ref. \cite{3} that
\[
\mid\mid \delta e \mid\mid^2 = \frac{\delta c^2}{c}-\int^1_0 e^3\zeta
\Delta \zeta,
\]
where $\Delta$ is the Laplacian
$\Delta\zeta=g^{-1}\frac{d}{dt}(e^{-1}\frac{d}{dt}(e \zeta))$.
The diffeomorphism $\zeta$ must obey the boundary conditions
$\zeta(0)=\zeta(1)=0$ for open paths and $\zeta(t)$
periodic for closed paths (the diffeomorphisms of [0,1] and $S_1$
respectively).

For closed paths, $\zeta=$ constant corresponding to global rotations
introduces a zero mode of $\Delta$. This will imply, as shown below, that
different
Jacobians occur for closed paths and open paths. The normalization for the
$\zeta$ integral is
\[
\int [d\zeta]\exp(-\frac{1}{2} \mid\mid \zeta \mid\mid^2)=1,
\]
where the invariant norm is
\[
\mid\mid \zeta \mid\mid^2 = \int^1_0 e^3 \zeta^2dt.
\]
since $\zeta$ transforms as a vector.

We  integrate over $\delta c$
and $\zeta$ to obtain $J$. The $\delta c$ integral contributes $(2\pi
c)^\frac{1}{2}$.
The integration over $\zeta$ depends on the boundary conditions. For open
paths we Fourier expand $\zeta(t)=\sqrt{2}\sum a_n \sin(n\pi t)$, $n>0$.
Then we
obtain
\[
\prod_{n>0}\int da_n\left(\frac{c^3}{2\pi} \right)^{\frac{1}{2}}\exp
(-\frac{1}{2} a^2_n
\frac{n^2\pi^2}{c^2} c^3) =
[\det_a(-\frac{1}{c^2}\frac{d^2}{dt^2})]^{-\frac{1}{2}} ] \sim
c^{-\frac{1}{2}}.
\]
The determinant is easily evaluated by $\zeta$ function regularization (see
ref. \cite{3}).  Reparametrization invariance  has been
exploited here to choose the gauge $e=c$. The Jacobian $J$ is therefore a
constant for
open paths.

In the case of closed paths $\zeta$ is periodic so we can expand $\zeta=
\sum b_n \exp(2\pi ni)$. The zero mode $b_0$ has to be included so that we
find a contribution.
\[
\prod_n \int db_n \left(\frac{c^3}{2\pi}\right)^{\frac{1}{2}} \exp
(-\frac{1}{2}b_n^2\cdot
\frac{4n^2\pi^2}{c^2}\cdot
c^3)=[\det'_b(-\frac{1}{c^2}\frac{d^2}{dt^2})]^{-\frac{1}{2}}L
\left(\frac{c^3}{2\pi}
\right)^{\frac{1}{2}}
\]
where $L$ is a regulator for the $b_0$ integral. Using $\zeta$ function
regularization again we find $\det' b\sim c^2$ since $n<0$ modes also
contribute. In this case the Jacobian $J\sim c^{-1}$.

The original partition function (8) can now be re-expressed as
\[
Z_g = \int_0^\infty dc J V_D \int [dx_\mu] \exp(-S_b[x_\mu,c])
\]
where $V_D=\int \mid df \mid$ is a formal diffeomorphic volume factor which
depends on $c$ \cite{9}. An analogous problem arises in the string case
where
the volume factor depends on the moduli \cite{7}. As for the string case
\cite{6,7} we define the physical partition function as
\begin{align}
Z_{phys} &= \int \frac{[de][dx_\mu]}{V_D} \exp(-S_g) \\
&= \int_0^\infty dc J(c) \int [dx_\mu]\exp(-S_g[x_\mu,c]).
\end{align}
The appearance of the volume term $V_D$ can be traced to the choice of
normalization made. This point is discussed further below in section 3.
It is satisfying to note that (10) now concurs with the Feynman proper
time single particle formalism for a bosonic string theory \cite{8} where
$c$ is the "proper time'' variable. This was illustrated in refs \cite{3,10}
where the correct propogator was calculated. In addition, the Jacobian $J$
introduces
the required $c$ dependence of open and closed paths. For closed paths $c$
also
plays the role of the proper time in the Schwinger formalism for evaluating
determinants \cite{9}.

As an example of a sum over closed paths we calculate the effective action
for a boson in an external electromagnetic field. The action is modified to
include a
reparametrization invariant interaction with the external field potential
$A_\mu(x_\mu)$ so that
\[
Z_A=\int_0^\infty \frac{dc}{c} \int[dx_\mu]\exp(-S_g+\int_0^1 \dot{x}_\mu
A_\mu dt).
\]

We can re-rexpress $Z_A$ as
\[
Z_A=\int_0^\infty \frac{dc}{c} e^{-\frac{m^2}{2}c}\int d^dy_\mu<y,c>\mid
y,0>,
\]
where
\[
<y,c \mid y,0> = \int[dx_\mu ]\exp(-\int_0^c
(\frac{1}{2}\dot{x}^2_\mu+A_\mu \dot{x}_\mu)
dt),
\]
where $\tau=ct$ is the ``proper Euclidean time'' and all paths begin and
end at $x_\mu=y_\mu$ (11) describes the evolution of a quantum mechanical
system with Hamiltonian $H=\frac{1}{2} (p-A)^2$ over a time $c$. Thus $Z_A$
becomes
\begin{align}
Z_A &=\int_0^\infty \frac{dc}{c}e^{-\frac{m^2}{2}c} Tr (e^{-cH}) \\
&= Tr \int_0^\infty \frac{dc}{c} e^{-\frac{1}{2}((p-A)^2+m^2)c} \\
&= Tr \log ((p-A)^2+m^2).
\end{align}
$Z_A$ represents the interaction of a single bosonic particle with an
external field and therefore exponentiating we recover the standard result
for a
bosonic field i.e. $\exp(Z_A)=\det ((p-A)^2+~m^2)$.

In the case of a constant external field $A_\mu=\frac{1}{2}x_\nu
F_{\mu\nu}$, $F_{\mu\nu}$ constant, so that the trace can be explicitly
calculated to
give the Euler-Heisenberg effective Lagrangian \cite{9}. Alternatively we
can compute
(11) directly since the $x$ integration is Gaussian. This is performed in
the
appendix.
\section{The Quantum Equivalence of $S$ and $S_g$}
We now demonstrate that the physical partition function of (9) is also
equivalent to the path length partition function $Z$. This equivalence will
be shown by
adopting an explicit Feynman prescription for the einbein path integral. The
correctness of the physical partition function prescription of (9) will be
demonstrated. We
begin by stating the result
\[
\exp (-S[x_\mu]) = \int [dw]\exp (-S_g[x_\mu,e=w^2]),
\]
where the $w$ measure is normalized to
\[
\int [dw] \exp (-\frac{1}{2}m^2\int w^2dt)=1
\]
corresponding to $x_\mu=0$ . We can think of this result as a
generalised Ehrenfest Theorem in the sense the average of length scale
fluctuations gives the quantum mechanics result.
To prove this we note firstly the useful integral identity
\begin{equation}
\exp(-\sqrt{ab})=\left(\frac{2a}{\pi}\right)^{\frac{1}{2}} \int_0^\infty dw
\exp(-\frac{1}{2}
aw^2-\frac{1}{2}bw^{-2}),
\end{equation}
where $a,b>0$. This identity follows from
\[
\int_0^\infty \exp
(-\frac{1}{2}(x-\frac{\alpha}{x})^2)=\sqrt{\frac{\pi}{2}}, \alpha \geq 0,
\]
which can be shown by differentiating with respect to $\alpha$.

To prove this consider an arbitrary finite partioning $t_0,\dots,t_n$
of the interval $[0,1]$ with $t_0=0$, $t_n=1$. We then express the action
$S$ as the Riemann sum to find $(\Delta_i t=t_i-t_{i-1})$
\begin{equation}
\exp(-S)=\lim_{\Delta_it \to 0} \prod_{i=1}^n \exp (-m\Delta_i t
((\frac{\Delta_ix^\mu}{\Delta_i t})^2)^{\frac{1}{2}})
\end{equation}
with $\Delta_i x^\mu=x^\mu(t_i)-x^\mu(t_{i-1})$ and $\Delta_it=t_i-t_{i-1}$.
 Now using (14) with $a_i=m^2 \Delta_i t$, $b_i=(\Delta_i x^\mu)^2(\Delta_i
t)^-1$ (15) becomes
\begin{eqnarray*}
\lim_{\Delta_i t \to 0} \int \prod_{i=1}^n dw_i \left(\frac{2m^2\Delta_i
t}{\pi}\right)^{\frac{1}{2}}
\exp(-\frac{\Delta_i t}{2}((\frac{\Delta_i
x^\mu}{\Delta_it})^2w_i^{-2}+m^2w_i^2)) \\
= \int [dw] \exp (-S_g[x_\mu,e=w^2]),
\end{eqnarray*}
The path integral measure $[dw]$ has been
explicitly specified by a standard Feynmann prescription. Under a
reparametrization both $S_g$ and $S$ remain invariant and hence the measure
is also invariant.
This can be seen directly at the discrete level since a reparametrization
corresponds
to a repartitioning $\{t_i\} \to \{s_i\}$ of the interval $[0,1]$. The
integrand
and measure are then clearly invariant under the discrete form of (6).

We can now change variables to $c$ and $f(t)$ as before. The Jacobian for
the transformation is again computed by working in the tangent space. The
invariant norm for 
$\delta w$ is
\[
\mid\mid \delta w^2 \mid\mid = \int^1_0 \delta w^2 dt = \frac{1}{4} \mid\mid
\delta e\mid\mid^2,
\]
with normalization
\[
\int d[\delta w] \exp(-\frac{1}{2}\mid\mid \delta w \mid \mid^2) =
\frac{k}{V_D},
\]
where $k$ is some constant for consistency with the earlier normalizations.

Defining the Jacobian $J_w$ by $[d(\delta w)]=J_w d(\delta c)[df]$
we find, by an argument similar to that above, that $J_w \sim J(c)/V_D$.
Transforming to the variables $c$ and $f$  we obtain
\[
Z \simeq \int_0^\infty \frac{dc}{c} \frac{J(c)}{V_D} \int [df]\int[dx]\exp
(-S_g) = Z_{phys}.
\]
Therefore the path length partition function $Z$ is equivalent to the
physical Polyakov-like partition function. It is interesting to note that
the normalization used is automatically consistent with the physical
definition
of the Polyakov partition function . This suggests that the natural
definition for the normalization of the tangent space measure $[d(\delta
e)]$
should be  the one used.

Finally, we note that since the invariant norm $\mid\mid \delta w \mid\mid$
and $\mid\mid \delta e \mid \mid$ are proportional, the Jacobian $J(c)$ and
$V_pJ_w$ must
also be proportional. Likewise, had we defined $Z_g$ as a sum over all
metrics
$g(t)$ then, since $\mid\mid \delta g \mid\mid=2\mid\mid \delta e \mid\mid$,
we again find the
same Jacobian in transforming to $c$ and $f(t)$.

\appendix

\section{Euler-Heisenberg Effective Action}

In this appendix we will calculate the effective action  in the
case where $F_{\mu\nu}$ is constant so that $A_\mu =\frac{1}{2}x_\mu
F_{\mu\nu}$.
We begin by Fourier expanding the periodic coordinate as $x_\mu=\sum
a^\mu_\nu
\exp(2n\pi ti)$. The action becomes (with $e=c$)
\[
S=\frac{1}{2} m^2c + \sum_n (\frac{2n^2\pi^2}{c}a^n_\mu a^{-n}_\mu + n\pi_i
a^{-n}_\mu
a^n_\nu F_{\mu\nu}).
\]
We can now calculate $Z_A$ by expanding in $F_{\mu\nu}$ using the Feynman
rules:
Propagator: $\frac{c}{4n^2\pi^2}$,
Vertex: $\delta_{\mu\nu}F_{\mu\nu}{n\pi}^2$
 We find that only even powers of $F$ contribute. In general
to $O(F^{2r})$  only one connected diagram  contributes
\[
\frac{1}{(2r)!}2^{2r-1}(2r-1)! Tr (F^{2r}) \sum_n
(\frac{c}{4\pi^2n^2})^{2r}(n\pi)^{2r}
\]
relative to the $O(F^0)$ contribution. The factors are respectively, a
symmetrizing factor for $2r$ identical $F$ sources, a combinatorial factor
for the number of
ways of connecting $2r$ vertices, a Lorentz index trace and a momentum sum
over
propagators and vertices. For simplicity we assume a constant electric field
only so
that $Tr(F^{2r})= 2(-1)^r E{2r}$. Summing over all connected diagrams we
find
\[
\sum_{r=1}^\infty \frac{1}{r} \left( \frac{Ec}{2\pi}\right)^{2r} \zeta (2r)
= \log \left(\frac{Ec/2}{\sin Ec/2} \right)
\]
where $\zeta(r)$ is the Riemann zeta function. We have used the
relation
\[
\sum_{r=1}^\infty (\frac{x}{\pi})^{2r} \zeta (2r) = \frac{1}{2}(1-x\cot x).
\]

The contribution from all disconnected diagrams can now be found by
exponentiating  the expression for the connected graphs .
Therefore the partition function  gives us
\[
Z_{EH} = \int_0^\infty \frac{dc}{c} \left( \frac{Ec/2}{\sin (Ec/2)}-1
\right) e^{-\frac{m^2c}{2}}
\int dy_\mu < y,c \mid y,0>_0
\]
where
\begin{align*}
<y,c \mid y,0>_0 &= \int [dx_\mu]\exp (-\frac{1}{2} \int_0^c \dot{x_\mu}^2
dt) \\
&=(2\pi c)^{-d/2},
\end{align*}
where all paths begin and end at $x_\mu=y_\mu$. Therefore we find that
$Z_{EH}$ is (with $c=2s$ and $L^d$ the volume of space)
\[
Z_{EH}=\frac{L^d}{(4\pi)^{d/2}} \int_0^\infty
\frac{ds}{s^{1+d/2}}(\frac{Es}{\sin Es}-1)e^{-sm^2},
\]
which is the standard result for a boson \cite{8}. The fermionic result is
easily found by also including a spin term which contributes an extra factor
to the
Lagrangian of $\frac{1}{4}F_{\mu\nu}\sigma_{\mu\nu}$. Tracing over the spin
we obtain the fermionic result \cite{8}.

\end{document}